# QUENCH PROTECTION ANALYSIS IN ACCELERATOR MAGNETS, A REVIEW OF THE TOOLS

H. Felice, LBNL, Berkeley, CA 94720, USA
E. Todesco, CERN, Geneva, Switzerland

*Abstract*

As accelerator magnets see the increase of their magnetic field and stored energy, quench protection becomes a critical part of the magnet design. Due to the complexity of the quench phenomenon interweaving magnetic, electrical and thermal analysis, the use of numerical codes is a key component of the process. In that respect, we propose here a review of several tools commonly used in the magnet design community.

## INTRODUCTION

Understanding the way the quench develops in the magnet and designing the appropriate protection scheme are key components of present accelerator magnet design. From the quench initiation and detection to the firing and connection of the protection elements, an ideal code would need to couple magnetic, electrical and thermal analysis. Nevertheless, the complexity of the topic leads to simplification in most of the codes commonly used. The disadvantage of a code containing all the physics is that computational times become long, not allowing parametric analysis. A common technique is to slice the problem, i.e. relying on different codes for different physical phenomena, and interfacing them.

In general, the user should choose the code depending on the most relevant physical phenomena she/he wants to model. Important added values of a code lies in (i) its ability to easily implement the geometry and the parameters of the magnet, (ii) the computational speed, (iii) the transparency of its contents, of the implemented physics and related approximations.

Even if not exhaustive, the review of the codes proposed in this paper is an attempt to provide some insight to the reader. We choose to distinguish two main categories of codes: the first one can be identified as "highly specialized" codes, dedicated to quench analysis, the second ones covers examples of Finite Element (FE) codes applied to quench protection analysis such as ANSYS or Cast3M.

## CODES CLASSIFICATION

*Physics*

The simplest level of modelling is Adiabatic Model, i.e. the implementation of the well-known equation that gives the balance between Joule heating due to the current and temperature increase through specific heats, ignoring the propagation of the heat through the coil. Since the specific heats and the resistivity have a pretty complex dependence on the temperature, the equation needs to be numerically integrated. An input parameter is the fraction of the coil which is in a resistive state; a second input parameter can be the quench velocity, allowing increasing the fraction of resistive coil with time to model quench propagation. QuenchPro are examples of adiabatic codes.

The second level of approximation is to include the Heat Propagation within the coil. According to the complexity of the geometry, this implies having a multidimensional mesh of the coil and insulation (eventually including ends). The code has to incorporate the magnetic field distribution in the coil and to include a model of the critical surface. It can therefore estimate the quench velocity, which becomes an output parameter.

The third level of approximation is to include also the Heat Exchange between the coil and the helium bath. This aspect is very relevant for fusion, where the cable in conduit is designed to have a large heat removal through the helium bath. For the impregnated coils of $Nb_3Sn$ this aspect is absent, whereas for the case of Nb-Ti coils permeated by helium, this aspect can play a more important role. In most cases for accelerator magnets, the heat exchange is ignored, and this approximation is considered to be conservative.

*Geometry*

The geometry of the coil is pretty complex, and trying a full modelling at the level of the cable plus insulation, or even for strand plus impregnation/voids and insulation may lead to prohibitive computational times. So it is wise to go by successive approximations. The origin of the complexity of the problem is (i) the large range of magnetic field in different parts of the coil, producing large differences in copper resistivity at low temperature, and (ii) the large variation of the specific heats with temperatures in the metal and in the resins.

The hierarchy of geometry modelling can be given as follows, from the simplest to the most complex level.

- Model the coil as a whole, with an average magnetic field and homogenized material properties.
- Model the coil having different layers, each one with its magnetic field.
- Model at the level of cables, each one with its magnetic field. The transposition of cables provokes an averaging over the different strands within the same cable.
- In principle one can think about reaching the level of strands: in this last case the void, insulation, and helium are modelled separately.

As far as we know, the modelling at the level of strands has never been done for accelerator magnets. On the other hand, this is done for cable in conduit used in fusion, where the complexity of the coil is not needed, so the model is just one cable, and the heat exchange between strands and between strand and helium is relevant.

If the model relies on a finite element module, the geometry is a mesh at the level of coil, layer, cable and (in principle) strand. The model can be an extrusion of the two dimensional cross-section, ignoring ends, or include the geometry of the ends. For the estimate of the hotspot temperature including the ends brings an additional complexity without adding anything relevant.

*Magnetic field and inductance*

The map of the magnetic field on the components/nodes of the geometrical model can be computed externally, i. e. is an input, or within a magnetic module of the code. This does not make a significant difference, except from the point of view of the computational time, the second option being more user friendly but much heavier.

What can be relevant in the case of very high field magnets with iron close to the coils is the impact of saturation. If the field map is simply scaled with the current the model is linear and the saturation is neglected. Otherwise, if the code has an internal algorithm to estimate the saturation, at each time iteration the field map is estimated (as in ANSYS or in ROXIE).

In case of significant saturation, the most important effect is its impact on inductance, which becomes dependent on the current. At each integration step one has to use the differential inductance to estimate the current decay. This effect can be estimated with a separate code, so that the inductance is a nonlinear known function of the current (as in the latest version of QuenchPro), or internally estimated through the code itself (as in ROXIE).

*Circuit*

All codes are usually able to simulate a circuit with the coil and a dump resistor. The current decays with instantaneous time constant $L/R(t)$, where $L$ is the differential inductance of the magnet and $R$ is the total resistance of the circuit, i.e. coil plus dump resistor. The resistance strongly depends on time, since the coil is heating. The inductance also depends on time if the saturation is non-negligible, since the current decays with time (see previous subsection).

A few codes are able to simulate a magnet made by several coils independently powered (as for instance QLASA). In this case one has several circuits coupled through the mutual inductances.

*Quench heaters*

For long and high field accelerator magnets, as the Nb-Ti main LHC magnets or the $Nb_3Sn$ HL-LHC triplet, the development of resistance due to the propagation of the spontaneous quench is negligible for estimating the hotspot temperature. On the other hand, the main mechanism is the development and propagation of the quench induced by the heaters. Several codes (QuenchPro, QLASA, …) are simulating it by initiating secondary quenches. A more complete simulation should include the heat propagation from the quench heaters to the coils. This problem can be treated separately through integration of heat equations, knowing the thermal properties and the critical surface; a network model has shown good results allowing modelling this complex phenomenon without free parameters [1]. Up to day this approach has not been fully integrated in the standard quench codes for superconducting accelerator magnets but the results of the simulation are used as inputs of other codes such as QLASA, QuenchPro or ROXIE.

## CODES DEDICATED TO SUPERCONDUCTING MAGNETS

Various codes addressing similar or complementary aspects of quench protections in superconducting magnets for accelerators have been developed in the past years. We propose here a summary of some codes frequently used by the accelerator magnets community.

*QuenchPro*

The QuenchPro is an adiabatic code where the coil is modelled at the level of layers. Coil ends are ignored. The field and inductance treatment is linear, but a nonlinear inductance has been introduced recently. Up to 16 layers can be modelled, but not with independent powering. The ouput provides two main parts:
- The first one is dedicated to temperature and resistance growth, hotspot temperature estimate and current decay computation.
- The second part focuses on voltages computation.

The magnet is subdivided in 16 sub-coils powered in series. The user has to provide the longitudinal quench propagation velocity for each layer; there is no propagation from turn to turn. Moreover, the user has the possibility to define the protection heater coverage, the value of the dump resistor and various delay times representative of the actual magnet protection scheme such as protection heater delay, quench detection time, validation window, switch time and so on. The material properties are homogenised over the cross-section of the conductor.

In the second part, the coordinates of all the turns in the magnet are defined allowing the computation of a turn-to-turn inductance matrix. This matrix along with the first part results (temperature, resistance and current decay) is then used to compute turn-to-turn and turn to ground voltage.

The program is presently written as a Mathcad spreadsheet. It has been developed at Fermilab by Pierre Bauer [2] and is presently available upon request to users like Giorgio Ambrosio at FNAL (giorgioa@fnal.gov).

*QLASA*

The QLASA is also an adiabatic code where the coil is modelled at the level of layers. The program has been developed at LASA [3]; programmed in Fortran 77, it was initially intended for quench analysis in superconducting solenoids and inductively coupled elements.

Despite the fact that the geometry is based on solenoids, it can be adapted to dipoles/quadrupoles with proper care requiring some discussions with the code developer (Massimo Sorbi) [3]. Ends are ignored, and field and inductance are treated linearly. Independent powering of different coils can be modelled.

The analysis is relying on an extensive material properties database MATPRO [4] and the default configuration of the program computes the quench propagation velocity based on these properties and the analytical formula. Nevertheless, the user has the flexibility to choose among several quench propagation velocity models or to input the quench velocity. As in QuenchPro, the program does not include thermal analysis and protection heaters are simulated using their delay as variable.

A strong feature of this code is to address the possibility of powering various layers with different power supplies or even without power supply in persistent mode.

### ROXIE Quench Module

The ROXIE Quench Module [5, 6] is an addition to the existing field computation program ROXIE developed at CERN [7], modelling the heat propagation within the coil. The geometry relies on the ROXIE input, at the level of cables. Then the straight section is discretized, ignoring the ends. The field map is estimated internally at each time step, including saturation if needed. The quench module uses a thermal network with a node per cable to solve the heat equation. The insulation between cables is also modelled with one node. From an electrical view point, the model assumes that the magnet is connected to an electrical circuit made of current source, diodes, extraction resistance. The critical surface of the superconductor, as the coil geometry and the magnetic field, is already available from the main ROXIE module.

The strength of this approach comes from the fact that it is integrated with the existing software, making many features and outputs from ROXIE available in the quench module with minimal work from the user. An example is the differential inductance of the magnet which is used in the electrical network. Another example is the use of coupling current losses in the heat equation as a heating factor of the conductor before quenching (called quench-back).

On the other hand, computational times require keeping the discretization to a limited number of nodes, which is not always enough for proper convergence. Therefore in some cases the user has to rescale some physical properties (as thermal conductivity) to match the measured values of the quench velocity.

Details on the numerical approach can be found in [5] and details on material properties are described in [6]. Some outputs of the code include voltage distribution, hot spot temperature and current decay.

Even if the core of the quench module is developed, and distributed as part of the ROXIE package, some developments are still being performed in particular regarding the modelling of the protection heaters which for now relies on a scaling factor tuning the heat transfer from the protection heaters to the conductor [8].

### Quench Analysis Program of Vector Field

The Quench Analysis Program is one of the analysis programs of the Opera-3D Analysis Environment from the Cobham Vector Field software [9]. It is a code that accounts for heat exchange in the coil (HE), based on the TEMPO/ Transient Thermal Analysis solver. It is coupled with the analysis of the electrical circuits and to the estimate of the fields in the conductors. The geometry is at the level of the cables (CA) and relies in the Vector Field input file. Homogenized material properties are used and need to be provided as part of the model creation.

In addition to the hot spot temperature and resistance growth computation, an interesting feature of the program is that it can be coupled with the ELEKTRA Time Varying Analysis to model transient electromagnetic fields and external circuits. This can be particularly useful in simulating transient eddy current in external magnet structure in thermal contact with the coils commonly referred as quench-back phenomenon. Despite these nice features, the software does not provide voltage computation. Voltage development being an important concern for long magnets, output post processing by the users is necessary.

Presently, the code is being used to perform the quench analysis of the MICE spectrometer and shows good agreement with test data [10]. Results are still to be published.

Regarding application to accelerator magnets, based on the experience of the authors so far, computation time remains very large which could be a showstopper.

## GENERAL CODES

In addition to the specialized codes, the use of commercial finite element codes is another way to model magnets and perform quench protection analysis. In both cases the codes model the heat propagation through the coil. Once the geometry is modelled, the program contains modules which account for the physics of the problem.

### Cast3M

Cast3M is a FEM code developed at CEA [11] and available at [12]. It is used at CEA/Saclay to perform eddy currents computation. An example is the case of CMS (Compact Muon Solenoid), where eddy currents were computed in the so-called quench-back cylinder [13, 14].

*ANSYS and other commercial codes*

ANSYS is a well-known FEM software available commercially. Some past work performed by Yamada et al. [15] and S. Caspi and P. Ferracin [16, 17] shows the advantage of such multi-physics models. In particular, with the increasing hot spot temperature in very demanding system, the capability to couple thermal and mechanical analysis is very interesting.

## CONCLUSION

This paper presents some codes commonly used in the community to perform quench protection analysis. The list is not exhaustive and many laboratories have their own "in-house" codes. In addition, considering quench protection codes used in the fusion community would also be highly beneficial. The attempt to collect more information about these various codes is still in progress.

## ACKNOWLEDGMENT

The authors would like to thank Giorgio Ambrosio, Berhnard Auchmann, Philippe Fazilleau, Massimo Sorbi, Heng Pan, Lidia Rossi, Tiina Salmi for their help.